\documentclass[12pt]{article}
\usepackage{amssymb}
\usepackage{epsf}
\usepackage{a4}

\newcommand{\btb}{\bar{\bt}}
\newcommand{\lmb}{\bar{\lm}}

\newcommand{\vb}{\bar{v}}

\newcommand{\vecx}{{\bf x}}       

\newcommand{\al}{\alpha}
\newcommand{\bt}{\beta}
\newcommand{\gm}{\gamma}
\newcommand{\dl}{\delta}
\newcommand{\ep}{\epsilon}

\newcommand{\et}{\eta}
\newcommand{\kp}{\kappa}
\newcommand{\lm}{\lambda}
\newcommand{\rh}{\rho}

\newcommand{\ta}{\tau}

\newcommand{\vr}{\varphi}

\newcommand{\om}{\omega}
\newcommand{\Om}{\Omega}

\newcommand{\Gm}{\Gamma}
\newcommand{\Lm}{\Lambda}
\newcommand{\Sg}{\Sigma}

\newcommand{\half}{\frac{1}{2}}

\newcommand{\Tr}{\mbox{Tr}\,}  

\newcommand{\dmu}{\partial_{\mu}}

\newcommand{\dnot}{\partial_{0}}

\newcommand{\eela}[1]{\label{#1}\end{equation}}
\newcommand{\eeala}[1]{\label{#1}\end{eqnarray}}
\newcommand{\ra}{\rightarrow}

\newcommand{\be}{\begin{equation}}
\newcommand{\ee}{\end{equation}}
\newcommand{\bea}{\begin{eqnarray}}
\newcommand{\eea}{\end{eqnarray}}
\begin{document}

\begin{titlepage}
\baselineskip=12pt
\rightline{ITFA-96-11}
\rightline{hep-ph/9605016}
\vskip .4in

\baselineskip=21pt

\begin{center}
{\Large{\bf 
Chern-Simons Diffusion Rate near the Electroweak Phase Transition 
for $m_H \approx m_W$
}}
\end{center}
\vskip .1in

\begin{center}
{\large Wai Hung Tang\footnote{email: tang@phys.uva.nl}\mbox{ } 
and
Jan Smit\footnote{email: jsmit@phys.uva.nl}\mbox{ } 
}

{\it Institute of Theoretical Physics, University of Amsterdam\\
Valckenierstraat 65, 1018 XE Amsterdam, the Netherlands}

May 9, 1996

\end{center}

\vskip 0.7in

\centerline{ {\bf Abstract} }
\baselineskip=20pt
\vskip 0.5truecm
The rate of $B$-violation in the standard model at finite 
temperature is closely related to the diffusion rate $\Gamma$ of 
Chern-Simons number. We compute this rate for $m_H \approx m_W$
in the classical approximation in an effective SU(2)-Higgs model, 
using Krasnitz's algorithm. The parameters in the effective 
hamiltonian are determined by comparison with dimensional 
reduction. In the high temperature phase we find $\Gamma/V (\alpha_W 
T)^4 \approx 1$, neglecting a finite renormalization. 
In the low temperature phase near the transition 
we find the rate to be much larger than might be expected from 
previous analytic calculations based on the sphaleron.
\end{titlepage}

\baselineskip=16pt

\section{Introduction}
Fermion number is not conserved in the standard model of 
elementary particle interactions, because of the anomaly in the 
divergence of the $B+L$ current. This leads to a rate $\Gm$ of 
fermion number violation which plays an important role in 
theories of baryogenesis \cite{CoKaNe93,RuSha96}. In the approximation 
of taking into account only the SU(2) contribution to the 
anomaly, $\Gm$ can be identified from the diffusive behavior of 
the `topological susceptibility' at large (real) times 
\cite{KleSha},
\bea
\Gm t &=& \langle Q^2(t) \rangle_T,
\;\;\; t\ra\infty, \label{1}\\
Q(t) &=& \int_0^t dx^0\, \int d^3 x\, q(x),\;\;\;
q = \frac{1}{32\pi^2}\, F_{\mu\nu}^{\al} \tilde{F}^{\mu\nu\al}.
\eea
Here $F$ is the SU(2) field strength and the bracket in (\ref{1}) 
denotes the expectation value at temperature $T$.

For temperatures below the electroweak phase transition, analytic 
calculations based on the sphaleron solution lead an expression 
which may be summarized in the form 
\cite{ArMcLe87,Caea90,BaJu94}
\bea
\frac{\Gm}{V} &=& \kp (\al_W T)^4,\label{kappa}\\
\kp &=& f(\frac{\lm_3}{g_3^2})\,    
\left(\frac{E_s(T)}{T}\right)^7\, \exp(-\frac{E_s(T)}{T}),
\label{Gan}
\eea
where $V$ is the volume, $\al_W = g^2/4\pi$ is the SU(2) gauge 
coupling, $\lm_3/g_3^2 \approx \lm/g^2 \approx m_H^2/8m_W^2$ is a 
coupling ratio in the 3D dimensionally reduced theory and 
$E_s(T)$ is the temperature dependent sphaleron energy.
The effect of the fermions enters mainly in the relation of the 
3D couplings to the original 4D standard model \cite{Moo95} 
(see also \cite{Diaea94}).

For high temperatures above the phase transition, 
dimensional reasoning leads to the form (\ref{kappa}) with 
$\kp$ roughly temperature independent \cite{CoKaNe93,RuSha96}
(a logarithmic dependence may still be present). An analytic 
estimate $\kp \approx 0.01$ was carried out in \cite{Ph95}. 

The rate $\Gm$ defined by (\ref{1}) is a nonperturbative quantity 
and it is important to check analytic calculations  by numerical 
simulation. This is well-known to be difficult since the 
effective Boltzmann factor in real time processes is complex:
in the temporal gauge $A_0 = 0$ we can write
\bea
Q(t) &=& N_{\rm CS}(t) - N_{\rm CS}(0), \\
\langle Q^2(t) \rangle_T
&=& \Tr e^{-H/T}\, \left[ e^{iHt}\, N_{\rm CS}(0)\, e^{-iHt} - 
N_{\rm CS}(0)\right]^2/\Tr e^{-H/T}, \label{2} 
\eea
where $N_{\rm CS}$ is the Chern-Simons number and the trace is over states 
in Hilbert space satisfying the Gauss constraint.
To cope with the complex weights a classical approximation has 
been  introduced in \cite{GrRu} and tested on the abelian-Higgs 
model in 1+1 dimensions \cite{GrRuSha}, using the microcanonical 
ensemble. Subsequent computations used Langevin methods 
\cite{BochFo93,KraPo94} and the canonical ensemble 
\cite{SmTa94,FoKraPo94,SmTa95}. In these computations the quantum mechanical 
expectation value (\ref{2}) is replaced by a classical 
expression,
\bea
\langle Q^2(t) \rangle_T
&=& Z^{-1} \int D\pi D\vr\, \dl(G)\, e^{-H_{\rm eff} 
(\vr,\pi)/T}\, \left[ N_{\rm CS}(\vr(t),\pi(t)) - N_{\rm CS}(\vr,\pi)\right]^2, 
\label{3}\\
Z &=& \int D\vr D\pi\, \dl(G)\, e^{-H_{\rm eff}(\vr,\pi)/T},
\eea
where $\vr$ and $\pi$ denote generic canonical variables and 
$\dl(G)$ enforces the classical Gauss constraint. 
Furthermore, $\vr(t)$ and $\pi(t)$ are solutions of Hamilton's 
equations with hamiltonian $H_{\rm eff}$ and initial conditions 
$\vr(0)=\vr$, $\pi(0)=\pi$. The classical partition function is 
still a functional integral over all $\vr(\vecx)$ and $\pi(\vecx)$ and 
needs regularization. In the numerical simulations this is 
provided by a lattice. 

In the abelian-Higgs model the numerical results agree with the
analytical sphaleron-type calculations in the low temperature regime
\cite{SmTa94,FoKraPo94}. In the high temperature regime the rate
appears to follow the 1+1 dimensional analogue of (\ref{kappa}) with
temperature independent $\kp$, which depends, however, on the lattice
spacing \cite{SmTa95} (see also \cite{FoKraPo94}). As emphasized in
\cite{SmTa95}, such a lattice spacing dependence should be cancelled
by a corresponding dependence in the hamiltonian $H_{\rm eff}$, which
is to be an effective hamiltonian appropriate for the lattice used.

In the realistic 3+1 dimensional case of the SU(2)-Higgs model,
exploratory simulations have been carried out in \cite{Ambea}.  The 
incorporation of the Gauss constraint has been a source
of uncertainty in the nonabelian case. Recently, a solution to the problem
has been presented \cite{Kra95} and first results have
been obtained for $\kp$ in the pure SU(2) case \cite{AmKra95}. Here we
study the full SU(2)-Higgs case using the algorithm and implementation
of ref.\ \cite{Kra95}. As this work was being completed a paper
appeared describing simulations with a chemical potential for the 
Chern-Simons number, using a different implementation of the Gauss
law \cite{Moo96}.

In sect.~2 we discuss the classical approximation in general 
terms and emphasize the relation to dimensional reduction (DR). 
In sect.~3 we describe the determination of the parameters in the 
effective hamiltonian for the SU(2)-Higgs model by comparison 
with the dimensional reduction results of refs.\ 
\cite{Faea1,Faea2}. The connection with dimensional reduction is 
also discussed in refs.\ \cite{RuSha96,AmKra95,Boch93}. In 
sect.~4 we go through some conventions used in the numerical 
implementation and in sect.\ 5 we present results for the rate 
$\Gm$ in a temperature region around the phase transition. 
Sect.~6 contains our conclusion.
Some details are delegated to the appendices.

\section{Classical approximation and dimensional reduction}

A classical approximation is expected to be valid if  expectation 
values are dominated by states with high mode occupation numbers. 
This will occur for temperatures higher than any mode energy, 
which suggests introducing a cutoff $\Lm$ to keep these 
energies bounded from above. We can imagine integrating out the 
modes on spatial momentum scales larger than $\Lm$, leading to an 
effective action $S_{\rm eff}$ for the remaining variables. If 
this effective system is weakly coupled, such that a description 
in terms of `modes' applies, with mode energies $\leq \sqrt{({\rm 
mass})^2 + \Lm^2}$, we may expect classical behavior for $\Lm \ll 
T$.

For gauge theories a separation in low and high energy modes is 
notoriously difficult. However, sensible approaches do exist, for 
example real space renormalization group transformations in 
lattice formulations \cite{RG}, or methods as outlined in 
\cite{KeMaPa}.  

For definiteness we assume $\Lm$ to be represented by a  spatial 
cubic lattice with spacing $a=\pi/\Lm$. Keeping time continuous 
seems artificial and perhaps we should also coarse-grain
in time. For slow processes a derivative expansion may be 
adequate and keeping only two time derivatives we can go over to 
a canonical formalism with an effective hamiltonian $H_{\rm 
eff}(\vr,\pi)$. The procedure is to be applied to observables $O$ 
which are dominated by low energy modes which can be 
expressed in the effective variables in a practical way, $O \ra 
O_{\rm eff}(\vr,\pi)$.

Under weak (effective) coupling conditions this hamiltonian 
$H_{\rm eff}$ may then be approximated by a classical form in which 
the parameters depend on $\Lm$ and $T$.
To avoid strong coupling in a nonabelian gauge theory, the scale 
$\Lm$ should presumably not be too low.
The procedure thus leads first to an expression of the form 
(\ref{2}), but with $H\ra H_{\rm eff}$ on a spatial lattice.
This quantum mechanical expression is then further approximated 
by the classical expression (\ref{3}). 

In non-gauge theories the hamiltonian $H_{\rm eff}$ has the typical form 
\be
H_{\rm eff} = \sum_x\, \frac{z}{2}\, \pi(\vecx)^2 + W(\vr),
\ee
where $W$ includes the spatial gradients. Performing the gaussian 
integration over the canonical momenta $\pi$ in the partition 
function gives a reduced partition function which has a 
dimensional reduction form
\be
Z_{\rm DR} = \int D\vr\, \exp[-S_{\rm DR}(\vr)],\;\;\;
S_{\rm DR}(\vr) = W(\vr)/T.
\ee
Correspondingly, the static (equilibrium) expectation values 
in the classical effective theory have the dimensional reduction 
form. So we may identify the classical approximation of the 
effective theory with a dimensionally reduced theory 
on a lattice.

The dimensional reduction method can be formulated as a matching 
procedure \cite{Kaea96}, in which the parameters in a local DR 
action are calculated in perturbation theory to give the same 
physical results as the original 4D theory, approximately. The 
error in the approximation is partially due to the neglect of 
nonlocal terms which appear first at the two loop level 
\cite{JaPaPe96}. 

Now the cutoff $\Lm$ in the DR theory is interpreted as an 
ultraviolet cutoff, which is to be removed to infinity while 
tuning the DR parameters in the appropriate way. However, it 
seems undesirable to remove the regulator in the reduced theory 
completely, since this would imply tuning to a second order phase 
transition, with accompanying loss of free parameters due to 
universality. For example, in the SU(2)-Higgs theory
the ratio $m_H/m_W$ would be fixed at the critical point.
Such a problem does not enter in the approach based on 
integrating out high momentum modes, in which the cutoff is not 
be sent to infinity, which implies of course some nonlocality.

But this method has its own awkward features when $\Lm \ll T$. 
Then all modes up to $\Lm$ are highly excited and the lattice 
nature of the system will be ubiquitous. To obtain cutoff 
independent results under such conditions we may have to take 
into account the regularization dependence of the effective 
observables. But the calculation of the effective hamiltonian is 
already difficult and has to our knowledge not been done so far, 
let alone the obervables. 

However, the phenomenon of dimensional reduction suggests 
considerable freedom in the choice of $\Lm$, including $\Lm \gg 
T$, provided the observables are dominated by low momenta. Using 
such cutoffs $\Lm \gg T$, regularization artefacts get suppressed 
and the latticized classical form of the observables will suffice 
(taking into account necessary renormalizations).
Most importantly, we are lead to identify the potential energy 
part of the effective hamiltonian  with $S_{\rm DR}$. 

The initial conditions in (\ref{3}) now correspond to the DR 
action whose renormalization properties are well understood. For 
short times one expects the real time correlators to have similar 
renormalization properties because Hamilton's equations imply a
continuous dependence on the initial conditions. An example of
a one loop perturbative calculation in scalar field theory can
be found in ref.\ \cite{BoMcLeSmil95}.
For large times the dependence on initial conditions is 
governed by Lyapunov exponents, which should be renormalized 
quantities (i.e.\ of the order of a physical mass scale and not 
of order of the cutoff) in order to have renormalizability of the 
real time correlators. Evidence for this is provided in ref.\ 
\cite{Kra95}, where Lyapunov exponents were measured in SU(2) 
gauge theory. The difference between two neighboring initial 
conditions $a$ and $b$ was characterized by electric and magnetic 
`distances' $D_{E,M}$,
\be
D_E \propto \int d^3 x\, |E_{(a)}^2 - E_{(b)}^2|,\;\;\;
D_M \propto \int d^3 x\, |B_{(a)}^2 - B_{(b)}^2|
\ee
(in continuum notation).
Assuming ultraviolet divergences to cancel in these differences,
the Lyapunov exponents extracted from their exponential
time behavior may be expected to be renormalized quantities. In 
ref.\ \cite{Kra95} it was found\footnote{In the notation
introduced in (\ref{betaG}),(\ref{betabar}) the result in 
\cite{Kra95} was $a\lm = 1.20 /\btb$, where $a$ is the lattice 
distance and $\btb = 4/g^2 aT$.} that $D_E$ and $D_M$ lead to the 
same Lyapunov exponent $\lm = 0.30\, g^2 T$.

\section{The effective gauge-Higgs hamiltonian}

We now look more closely at the SU(2)-Higgs theory with classical 
action
\be
S = -\int d^4 x\, [\frac{1}{4g^2} F_{\mu\nu}^{\al} F^{\mu\nu\al} 
+ ((\dmu-iA_{\mu})\vr)^{\dagger} (\partial^{\mu} -iA^{\mu})\vr
+ \mu^2 \vr^{\dagger}\vr + \lm (\vr^{\dagger}\vr)^2],
\label{SU2HC}
\ee
where $\vr$ is Higgs doublet and
$A_{\mu} = A_{\mu}^{\al} \ta_{\al}/2$, with $\ta_{\al}$ 
the three Pauli matrices, $\al = 1,2,3$. Our basic assumption is 
that a reasonable approximation to the effective action is 
provided by a lattice version of (\ref{SU2HC}):
\bea
S_{\rm eff} &=& \int dx^0\, L_{\rm eff},\\
L_{\rm eff} &=& \sum_{\vecx}\left[
\frac{1}{z_E g_{\rm eff}^2}\, 
\Tr (D_0 U_{m\vecx})^{\dagger} D_0 U_{m\vecx} 
+ \frac{1}{z_{\pi}}\, 
((\dnot -iA_{0\vecx})\vr_{\vecx})^{\dagger} 
 (\dnot -iA_{0\vecx})\vr_{\vecx}
\right]
\nonumber\\ &&\mbox{} - W, \label{z}\\
W &=& \sum_{\vecx}
\left[\sum_{mn}\frac{1}{g_{\rm eff}^2}\,\Tr(1-U_{mn\vecx})
+ (D_m\vr_{\vecx})^{\dagger} D_m\vr_{\vecx}
\right.\nonumber\\ && \left. \mbox{}
+ \mu_{\rm eff}^2 \vr^{\dagger}_{\vecx} \vr_{\vecx}
+ \lm_{\rm eff} (\vr^{\dagger}_{\vecx} \vr_{\vecx})^2 +\ep
\right],
\label{5}\\
D_0 U_{m\vecx} &=& \dnot U_{m\vecx} - i A_{0\vecx} U_{m\vecx} + 
iU_{m\vecx} A_{0\vecx + \hat{m}}, \;\;\;
D_m\vr_{\vecx}= U_{m\vecx}\vr_{\vecx+\hat{m}} - \vr_{\vecx}, 
\eea
where we used lattice units $a=1$. The $U_{m \vecx} 
\leftrightarrow \exp(-iaA_{m\vecx})$ are the parallel 
transporters at $\vecx$ for directions $m=1,2,3$ and 
$U_{mn\vecx}$ is the transporter around a plaquette. 
The couplings $z_E$, $z_{\pi}$, $g_{\rm eff}^2$, $\mu_{\rm 
eff}^2$ and $\lm_{\rm eff}$ depend on the temperature, lattice 
spacing and the renormalized couplings $g^2$, $\mu^2$, $\lm$ in 
some renormalization scheme. We have kept the standard 
normalization for the spatial gradients, which defines $g_{\rm 
eff}^2$ 
and the other couplings. The kinetic terms have renormalization 
factors $z_E$ and $z_{\pi}$, which will be of the form $z_{E,\pi} 
= 1+ O(g^2,\lm)$. They influence the time scale relative to the 
momentum scale. The constant $\ep$ adjusts the energy density and 
plays no dynamical role in our theory (it takes care of the 
Rayleigh-Einstein-Jeans divergence). 

The action is invariant under gauge transformations 
$\Om_{\vecx}(x^0)$ acting in the usual way. A canonical 
description is straightforward in the temporal `gauge' $A_0=0$, 
leading to the hamiltonian
\be
H_{\rm eff} = \sum_{\vecx} \left(\frac{z_E}{2}\, g_{\rm eff}^2
E^{\al}_{m\vecx} E^{\al}_{m\vecx}
+ z_{\pi} \pi_{\vecx}^{\dagger} \pi_{\vecx}\right)
+ W,
\label{6}
\ee
where $E^{\al}_{m\vecx}$ is the left translator 
$E^{\al}_{m\vecx}(L)$ or right translator $E^{\al}_{m\vecx}(R)$ 
(cf.\ appendix A for more details), and $\pi_{\vecx}^{\dagger}$ is 
conjugate to $\vr_{\vecx}$. The residual invariance under static 
gauge transformations is dealt with by imposing the Gauss 
constraint. In the quantum theory $G_{\vecx}^{\al}|{\rm phys} \rangle =0$,
with
\bea
G_{\vecx}^{\al}&=&\frac{\dl S_{\rm eff}}{\dl A_{0\vecx}^{\al}}  
= (-D'_m E_{m\vecx}^{\al}+j^{0\al}_{\vecx}),\\ 
D'_m E_{m\vecx}^{\al}&=& \sum_m [E^{\al}_{m\vecx}(L)
- E^{\al}_{m\vecx-\hat{m}}(R)],\\
j^{0\al}_{\vecx}&=&  
i\vr_{\vecx}^{\dagger} \frac{\ta_{\al}}{2}\,\pi_{\vecx}
-i\pi^{\dagger}_{\vecx} \frac{\ta_{\al}}{2}\,\vr_{\vecx}.
\eea
In the classical approximation, the partition function entering 
in (\ref{3}) reads for the SU(2)-Higgs theory (cf.~appendix A)
\be
Z = \int DE D\pi DU D\vr\, [\prod_{\vecx\al} 
\dl(G_{\vecx}^{\al})] \, \exp(-H_{\rm eff}/T).
\label{8}
\ee
If we now write
\be
\dl(G_{\vecx}^{\al}) = \int \frac{dA_{0\vecx}^{\al}}{2\pi}\, 
\exp(iA_{0\vecx}^{\al} G_{\vecx}^{\al}),
\ee
carry out the integrations over $E$ and $\pi$ and subsequently 
rescale $A_{0\vecx}^{\al} \ra \sqrt{z_E} A_{0\vecx}^{\al}/T$, we 
arrive at the dimensional reduction form
\be
Z_{\rm DR} = \int DA_0 DU D\vr \exp(-S_{\rm DR}),
\ee
with
\bea
S_{\rm DR} &=&
\frac{W}{T} + \frac{1}{g_{\rm eff}^2 T} \sum_{\vecx} [\half\, D_m 
A_{0\vecx}^{\al} D_m A_{0\vecx}^{\al} + \frac{z_E/z_{\pi}}{4}\, 
g_{\rm eff}^2
\vr^{\dagger}_{\vecx}\vr_{\vecx} A_{0\vecx}^{\al} A_{0\vecx}^{\al}],
\label{7}\\
D_m A_{0\vecx}^{\al} &=& R_{\al\bt}(U_{m\vecx}) 
A_{0\vecx+\hat{m}}^{\bt} - A_{0\vecx}^{\al}
\eea
($R_{\al\bt}(U)$ is the adjoint representation of $U$).
This fits into the general DR form used in \cite{Faea1,Faea2} in 
numerical simulations,
\bea
S_{DR} &=& \bt_G \sum_{\vecx m<n} (1-\half\, \Tr U_{mn\vecx})
-\bt_H\sum_{\vecx m} \half\,\Tr\Phi^{\dagger}_{\vecx} U_{m\vecx}\Phi_{\vecx+\hat{m}} 
\nonumber\\ &&\mbox{}
+\sum_{\vecx} \half\, \Tr \Phi^{\dagger}_{\vecx} \Phi_{\vecx} 
+ \bt_R \sum_{\vecx} (\half\, \Tr \Phi^{\dagger}_{\vecx} \Phi_{\vecx})^2
\nonumber\\ &&\mbox{} 
+ \bt_G\sum_{\vecx m} \half\, \Tr (A_{0\vecx}^2 
- A_{0\vecx} U_{m\vecx}^{\dagger} A_{0\vecx+\hat{m}} U_{m\vecx})
+ \frac{\bt_H}{2}\sum_{\vecx} \half\,\Tr A_{0\vecx}^2 \half\Tr \Phi^{\dagger}_{\vecx} 
\Phi_{\vecx}
\nonumber\\ &&\mbox{} 
- \bt_2^A \sum_{\vecx} \half\, \Tr A_{0\vecx}^2 
+ \bt_4^A \sum_{\vecx} (\half\, \Tr A_{0\vecx}^2)^2,
\eea    
with
\bea
\frac{4}{g_{\rm eff}^2 aT} &=& \bt_G,\;\;\;
\frac{\lm_{\rm eff}}{g_{\rm eff}^2} = \frac{\bt_G \bt_R}{\bt_H^2},
\;\;\;
a^2 \mu_{\rm eff}^2 = \frac{2(1-2\bt_R-3\bt_H)}{\bt_H},
\label{btR}\\
\frac{z_E}{z_{\pi}} &=& 1,\;\;\;
\bt_2^A = \bt_4^A =0
\label{zrat}
\eea
(we have redisplayed the lattice distance $a$).
It is clear that in customary DR notation $g_{\rm eff}^2 T = 
g_3^2$, $\lm_{\rm eff} T = \lm_3$, for lattice spacing $a\ra 0$ 
in perturbation theory.

The parameters $\bt_G$ etc.\ are related to the 4D couplings, 
lattice spacing and temperature, depending on the 4D model 
\cite{Kaea96}. For the SU(2)-Higgs model\footnote{These formulas
taken from \cite{Faea1,Faea2} contain an error 
which is negligible for our purpose (cf.\
the comments below eqs.\ (142), (143) in \cite{Kaea96}).}
\bea
\frac{4}{g^2 aT} &=& \bt_G, \label{betaG}\\
\frac{m_H^2}{4T^2} &=& \left(\frac{g^2\bt_G}{4}\right)^2
\left\{ 3-\frac{1}{\bt_H} 
+ \frac{1}{\bt_G}\left[\frac{\rh\bt_H}{4} - \frac{9}{2} 
\left(1+\frac{\rh}{3}\right)\Sg\right] \right. \nonumber\\
&&\mbox{}
\left. - \half\, 
\left(\frac{9}{4\pi\bt_G}\right)^2\left[\left(1+\frac{2\rh}{9} - 
\frac{\rh^2}{27}\right) \ln \frac{g^2\bt_G}{2} + \et + 
\frac{2\rh}{9}\,\bar{\et} - \frac{\rh^2}{27}\,\tilde{\et}\right] 
\right\}
\nonumber\\
&&\mbox{}  
+ \frac{g^2}{2}\left[ \frac{3}{16} + \frac{\rh}{16} + 
\frac{g^2}{16\pi^2} \left(\frac{149}{96} + \frac{3\rh}{32}\right) 
\right],
\label{MH} \\
-\bt_2^A &\approx& \frac{5}{3}\, \frac{4}{g^2 \bt_G} 
-10\Sg,\;\;\;
\bt_4^A \approx \frac{17g^2}{48\pi^2}\, \bt_G,\\
\rh &=& \frac{m_H^2}{m_W^2},
\eea
with $\Sg = 0.252731$, $\et \approx 2.18$, $\bar{\et} \approx 1.01$,
$\tilde{\et} \approx 0.44$ \cite{Faea1,Faea2}. The 4D couplings 
$g^2$, $\lm$, $m_H^2$ and $m_W^2$ are $\overline{MS}$ scheme 
quantities at the scale $\mu_T =4\pi T e^{-\gm} \approx 7T$ 
\cite{Faea1}. The mass ratio
$m_H^2/m_W^2$ is approximately equal to its zero temperature 
value. We also have $\rh \approx 8\lm/g^2\approx 8\lm_{\rm 
eff}/g_{\rm eff}^2$, with small corrections of order $g^2$.

The $\bt_4^A$ term in the DR action is negligible in practice ($g\approx 
2/3$). The coefficient ratio $\bt_2^A/\bt_G$ is the Debye mass 
$m_D^2 = 5 g^2 T^2/12$ corrected with a counterterm $-5g^2 
T\Sg/12$. Setting $\bt_2^A$ to zero means that the Debye mass 
generated by (\ref{8},\ref{7}) will not have its standard 
perturbative value, but the accuracy of this perturbative value 
is unclear anyway. Our numerical results for the transition 
temperature to be described later indicate that the $\bt_{2,4}^A$ 
couplings in the effective action are not very important.

The relation $z_E/z_{\pi}=1$ is an approximation used in 
\cite{Faea2}; the corrections are small (cf.\ eq.\ (18) in 
\cite{Faea1}, $z_E/z_{\pi} = h_3/(g_3^2/4)$). With the 
approximation $z_E = z_{\pi}=z$, this parameter only affects the 
time scale. It takes care that the velocity of light $c=1$. In 
the following we shall absorb $z$ in the time scale
(i.e.\ set $z=1$),  which means that $c\neq 1$ in our units. 
However, we expect that $z = 1 + O(g^2,\lm)$ is close to 1 
anyway.

\section{Numerical implementation}
In the classical approximation the rate $\Gm$ can be computed by 
numerical simulation according to (\ref{3}). We use the algorithm 
and numerical implementation offered in ref.\ \cite{Kra95}. For 
clarity we record the conventions used in \cite{Kra95}, obtained by 
extracting an overall factor $4/g_{\rm eff}^2$ from the effective 
lagrangian (\ref{5}) and going over again to canonical variables 
(cf.\ appendix A). Indicating the quantities of ref.\ \cite{Kra95} 
by a bar we have
\bea
\frac{H_{\rm eff}}{T} &=& \btb \bar{H},\;\;\;\btb = \bt_G, \label{betabar}\\ 
\bar{H} &=& \sum_{\vecx} [\half\, \bar{E}^{\al}_{m\vecx} \bar{E}^{\al}_{m\vecx} 
+ \bar{\pi}^{\dagger}_{\vecx}\bar{\pi}_{\vecx} 
+ \sum_{m<n} (1-\half\, \Tr U_{mn\vecx})
\nonumber\\ &&\mbox{} 
+ (D_m\bar{\vr}_{\vecx})^{\dagger}D_m\bar{\vr}_{\vecx} 
+ \lmb(\bar{\vr}^{\dagger}_{\vecx}\bar{\vr}_{\vecx} - \vb^2)^2],
\eea
with
\be
\lmb = 4\frac{\lm_{\rm eff}}{g_{\rm eff}^2},\;\;\;
\vb^2 = - a^2\mu_{\rm eff}^2 \frac{g_{\rm eff}^2}{8\lm_{\rm 
eff}}.
\label{lmbvb}
\ee
Hamilton's equations follow from the nontrivial Poisson 
brackets
\be
\{U_{m\vecx},\bar{E}^{\al}_{m\vecx}(R)\} = iU_{m\vecx} \ta_{\al},
\;\;\;
\{\bar{\vr}_{\vecx},\bar{\pi}^{\dagger}_{\vecx}\} = 1.
\ee

Our investigation has concentrated on the case $m_H \approx m_W$. 
Using eq.\ (\ref{lmbvb}) we fix the parameter $\lmb=1/2$ since 
$2\lmb = 8\lm_{\rm eff}/g_{\rm eff}^2 \approx 8\lm/g^2 \approx 
m_H^2/m_W^2 = \rh = 1$. The temperature and lattice spacing 
dependence of $\lmb$ can be neglected. The quantities $\bt_R$ and 
$\vb^2$ are then given in terms of $\bt_H$ and $\bt_G$ by 
(\ref{btR}) and (\ref{lmbvb}), in particular
\be
\vb^2 = \frac{2}{\rh}\, 
\left(3 + \frac{\rh\bt_H}{\bt_G} - \frac{1}{\bt_H}\right).
\label{VB}
\ee
Eq.\ (\ref{MH}) gives an analytic constraint on the lattice 
spacing and temperature dependence of $\vb^2$. It is instructive 
to rewrite (\ref{MH}) in the form
\bea
4\lmb\vb^2 &=& a^2 m_H^2
- 
\left(\frac{3}{2}\,g\right)^2\,
\left(
\frac{3+\rh}{18} - \frac{3+\rh}{3}\, \frac{2\Sg}{aT}\, 
\right)\, a^2 T^2
- 
\left(\frac{3}{2}\,g\right)^4\,
\frac{1}{8\pi^2}\, \left( 
\frac{149+9\rh}{486}
\right.\nonumber\\
&&\left.
\mbox{}
 + \frac{27 + 6\rh -\rh^2}{27}\, 
\ln\frac{aT}{2} - \frac{27\et + 6\rh\bar{\et} - \rh^2 
\tilde{\et}}{27} \right)\, a^2 T^2,
\label{VBT}
\eea
which shows how $\vb^2$ depends on $a$ and $T$ if we neglect the 
temperature variation of $m_H$. The latter runs rather slowly 
with $T$ at $\mu_T \approx 7T$ and $T$ near $T_c$. We also 
neglect the running of $g^2$ and fix this coupling as $g=2/3$; 
then $aT = 9/\btb$. The temperature in units of $m_H$ now follows 
from (\ref{MH}) or (\ref{VBT}).

We also follow \cite{AmKra95} in the lattice implementation of 
the topological charge density,
\be
q_{\vecx}^L =  \frac{i}{32\pi^2}\sum_{k} 
[\bar{E}^{\al}_{k\vecx-\hat{k}}(R) + \bar{E}^{\al}_{k\vecx}(L)]\,
\Tr [(U_{l,m;\vecx} + U_{m,-l;\vecx} + U_{-l,-m;\vecx} + 
U_{-m,l;\vecx}) \ta_{\al}],
\label{qlat}
\ee
with $k,l,m=1,2,3 ({\rm cycl})$ (the sum is over identically 
oriented Wilson loops in the plane perpendicular to direction $k$ 
along the four plaquettes starting and ending at $\vecx$). The 
normalization of $q_{\vecx}^L$ is guided by the continuum limit 
$a\to 0$. The magnetic field strength $F_{lm}^{\al}(\vecx)$ is 
identified from any of the four plaquettes in (\ref{qlat}) in the 
usual way and the electric field strength is identified from 
$-2\bar{E}^{\al}_k(\vecx)\to \partial_0 A_k^{\al}(\vecx) = 
F_{0k}^{\al}$ by Hamilton's equations.  

In the context of zero temperature QCD this lattice version of 
$q(\vecx)$ is sometimes called a 
`naive' implementation (see \cite{Kro} for example for a review). 
As there is no other gauge invariant pseudoscalar field of 
dimension four to mix with, one expects that a finite 
multiplicative renormalization
is sufficient to obtain the desired observable,
\be
q(\vecx) = \kp_q q_{\vecx}^L.
\label{kpq}
\ee
This renormalization may be substantial. In perturbation theory 
$\kp_q$ is largely due to leaf (or `tadpole') diagrams, resulting 
from the compact nature of the lattice gauge fields, but the 
perturbative value is often unreliable. In the QCD context values 
for $\kp_q$ of order 5 have been reported in ref.\ \cite{DiGia} 
(where $\kp_q$ is denoted by $Z^{-1}$).
Ideally one would like to use fermions to measure directly how
they are affected by changes in the topology of the gauge field, 
as in the `fermionic approach' to topological charge in QCD 
\cite{SmVi,Vi}. For an interesting study of chirality and zero 
modes in the present context see ref.\ \cite{AmbFaHaKouThor}. We 
have not calculated $\kp_q$ yet.

\section{Numerical results}
Numerical results for equilibrium quantities using dimensional 
reduction have been obtained in ref.\ \cite{Faea2}. We use the 
same parameters at the phase transition as in this paper
(except that $\bt_{2,4}^A =0$, of course). Using the critical 
values $(\bt_G^c, \bt_H^c)$ at various lattice sizes $N^3$ from 
\cite{Faea2} gives $\vb_c^2$ according to eq. (\ref{VB}), shown 
in Table~\ref{v2table}.

\begin{table}
\centerline{\begin{tabular}{||c|r|l|l|l||} \hline
\multicolumn{1}{||c|}{$\bar{\beta}_c$} &
\multicolumn{1}{c|}{N} &
\multicolumn{1}{c|}{$\beta^c_H$} &
\multicolumn{1}{c|}{$\bar{v}^2_c$} &
\multicolumn{1}{c||}{$T_c/m_H$}
\\ \hline
12 & 8 & 0.3487 & 0.27894 & 1.91 \\ \hline
12 & 20 & 0.347733 & 0.26295 & 2.14 \\ \hline
12 & 24 & 0.34772 & 0.26273 & 2.15 \\ \hline
20 & 32 & 0.34173 & 0.15597 & 2.15 \\ \hline
\end{tabular}}
\caption{
Critical values of $(\btb,\vb)$ based on the DR data for
$(\bt_G^c,\bt_H^c)$ presented in ref.\  \protect \cite{Faea2}.
}
\label{v2table}
\end{table}  

\begin{figure}
\epsfysize 8.5cm
\centerline{\epsfbox{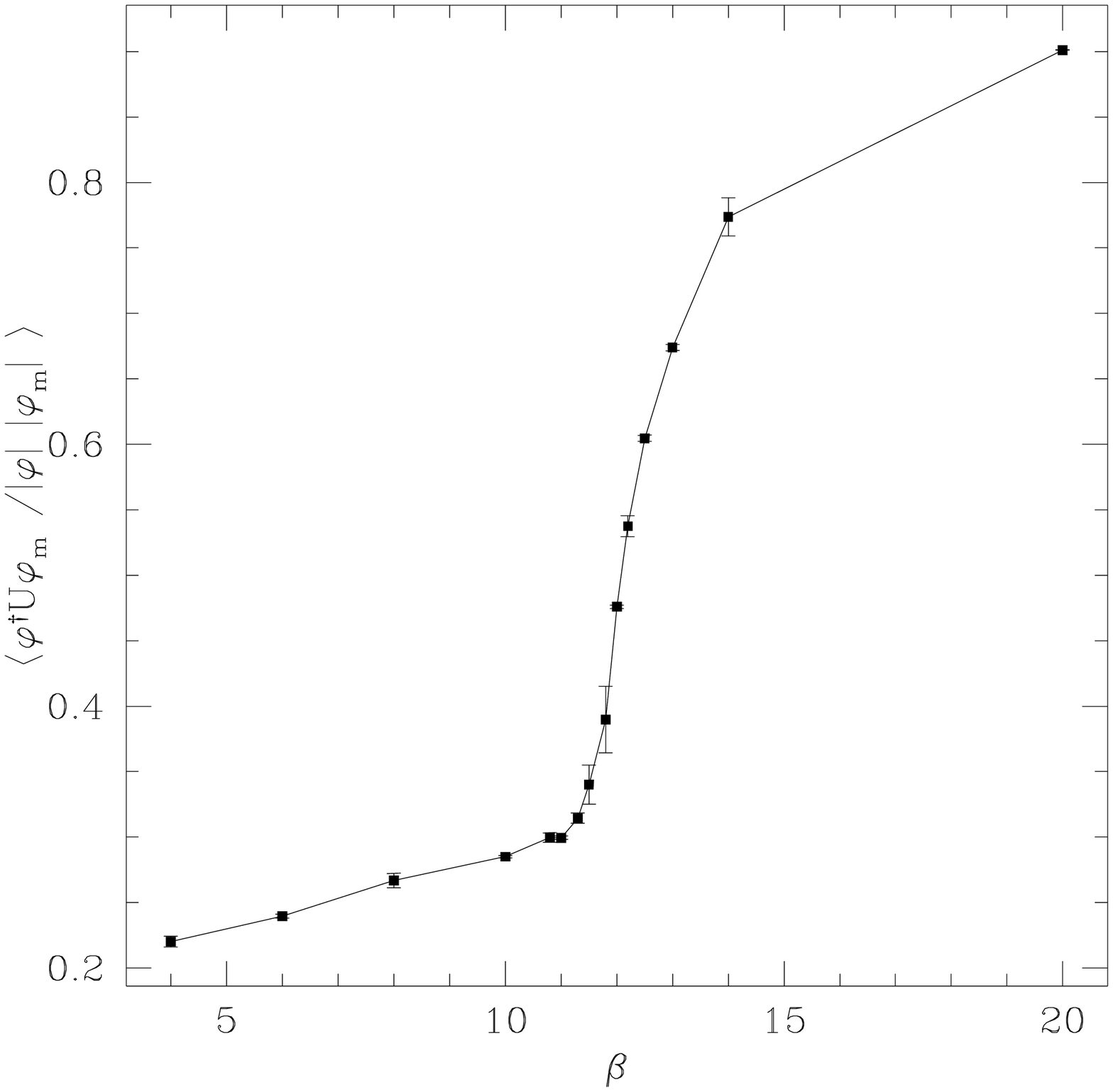}}
\caption{The correlation $\vr^{\dagger}_{\vecx} U_{m\vecx} 
\vr_{\vecx + \hat{m}}/|\vr_{\vecx}| |\vr_{\vecx + \hat{m}}|$ 
as a function of $\btb$ for $N=24$ at fixed 
$\vb^2=\vb^2_c = 0.263$.}
\label{f1}
\end{figure}

Since our effective hamiltonian implies vanishing couplings 
$\bt_{2,4}^A$ of the $A_0$ field in the associated DR theory, we 
want to see if this has an effect on the critical temperature.
In ref.\ \cite{Faea2} $\bt_G$ has been fixed while $\bt_H$ was 
varied to find the phase transition. Instead, we have kept 
$\vb^2$ fixed at $\vb^2_c$ and varied $\btb$. An example is given 
in Fig.~\ref{f1}, where $\vb_c^2$ corresponds to $\bt_G^c=12$ 
in \cite{Faea2}. We see that $\btb_c \approx 12 $  as well. From 
similar results for other observables at various volumes and 
pairs $(\bt_G^c,\bt_H^c)$ we have concluded that in all cases the 
critical $\btb_c$ was approximately equal to the input $\bt_G^c$ used 
for the calculation of $\vb^2_c$
(we have not made a precise determination of the transition 
values $\btb_c$, using e.g.finite size scaling). Hence, the fact 
that the effective hamiltonian implies $\bt_{2,4}^A =0$ does not 
seem to be important for the couplings used in our study, at 
least not for the value of $\btb_c$. In the following we set
$\btb_c=\bt_G^c$, as indicated in Table~\ref{v2table}.

Different values of $\btb_c$ correspond to different 
lattice spacings $a=4/g^2 T \btb$.
As a consistency check we may compare the $a$-dependence of 
$\vb_c^2$ obtained from the measured values of $\bt_H^c$ in ref.\ 
\cite{Faea2} with that predicted by eq.\ (\ref{VBT}). The two 
pairs $(\btb^c,N) = (12,20)$ and (20,32) are suitable for such a 
check as they correspond to approximately the same physical 
lattice size $L=Na$: $32 \approx (20/12)*20 = 33.3$.
Starting from the measured value of 
$\vb_c^2= 0.26295$ at 
$(\btb_c,N) = (12,20)$, eq.\ (\ref{VBT}) gives
$\vb_c^2 \approx 0.162$ at 
$(\btb_c,N) = (20,32)$ (using $g=2/3$ and $\rh = 1$), which is 
close to the measured value $\approx 0.156$
in Table~\ref{v2table}.
Neglecting finite size and additional lattice spacing effects, the
critical $\bar{\beta}_c$ and $\bar{v}_c^2$ values lead via (\ref{VBT})
to the ratios of $T_c/m_H$ shown in the right column of Table~\ref{v2table}. 

\begin{figure}
\epsfysize 8.5cm
\centerline{\epsfbox{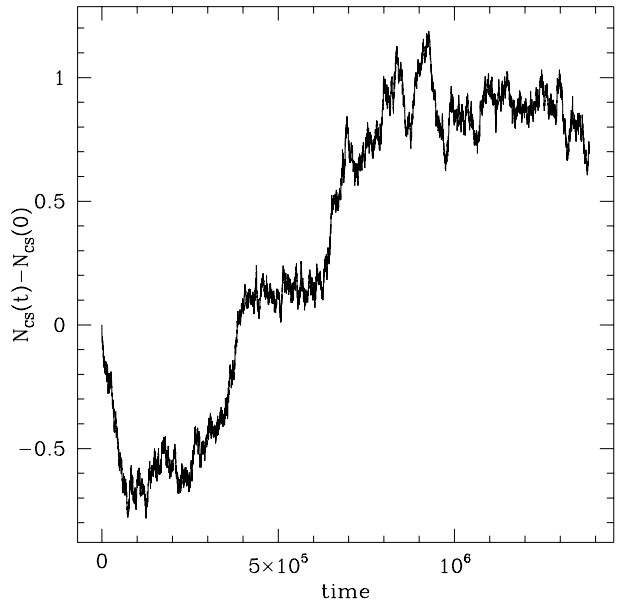}}
\caption{Hamilton evolution of $Q_L(t) = N^L_{\rm CS}(t) - 
N^L_{\rm CS}(0)$ on an $8^3$ lattice starting from an equilibrium 
configuration at $\btb=36$, $\vb^2 = 0.1564$.
}
\label{f2}
\end{figure}

Our results for the Chern-Simons diffusion rate do not 
incorporate yet the renormalization factor $\kp_q$. An 
illustration of the effect this can have is in Fig.~\ref{f2}, 
which shows the Hamilton evolution of $Q_L(t) = \int_0^t 
dx^0\, \sum_{\vecx} q_{\vecx}^L$ starting from a configuration 
equilibrated at $\btb=36$, with $\vb^2 = 0.1564$ corresponding 
to $\btb_c=12$, $\vb^2_c = 0.279$ via (\ref{VBT}). This is a 
rather large $\btb$ value corresponding to a low temperature
$T \approx T_c/3$ (neglecting finite size effects).
We see fluctuations about classical vacua separated by 
$\Delta N_{\rm CS} \approx 0.75$. With $\kp_q = 1/0.75 = 1.33$ we 
would get $\Delta N_{\rm CS}\approx 1$. Writing this as $\kp_q 
\approx 1 + 12/\btb$ would give $\kp_q \approx 2$ at $\btb=12$, 
which indicates that the $\kp_q$ correction can be substantial 
indeed.

The fact that 
the difference of Chern-Simons number between two adjacent vacua
in Fig.~\ref{f2} is near 1 also 
provides an indirect check on the normalization of the lattice 
form of $q_{\vecx}^L$.
This is relevant in view of the fact that the numerical rate 
turns out to be very different from expectations based on 
sphaleron calculations.

\begin{figure}
\epsfysize 8.5cm
\centerline{\epsfbox{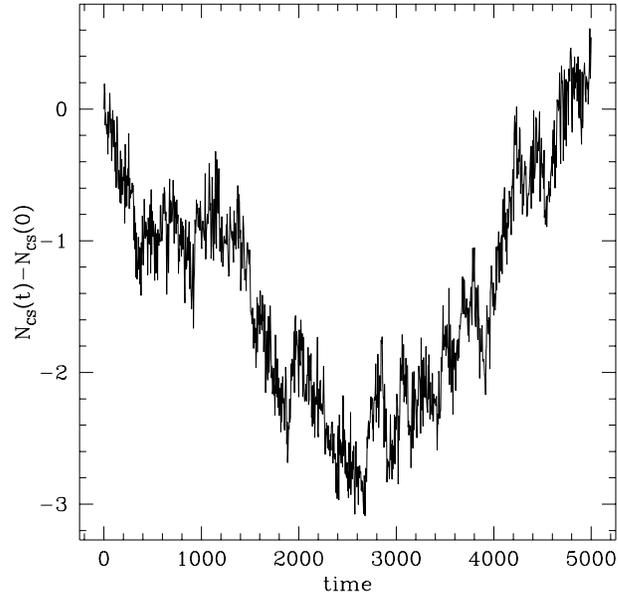}}
\caption{
Typical Hamilton evolution of $Q_L(t)$ starting with an equilibrium 
configuration at $\btb = 13$, $N=24$, $\vb^2 = \vb^2_c = 0.263$.
}
\label{f3}
\end{figure}

\begin{figure}
\epsfysize 8.5cm
\centerline{\epsfbox{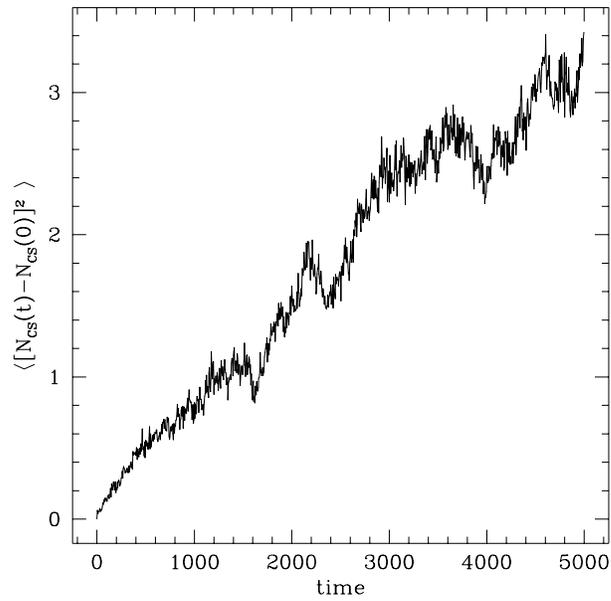}}
\caption{
The diffusion $\langle Q_L^2(t)\rangle$ 
for $\btb=13$, $N=24$, $\vb^2 = \vb^2_c = 0.263$, $t_{\rm run} = 
5000$.
}
\label{f4a}
\end{figure}

\begin{figure}
\epsfysize 8.5cm
\centerline{\epsfbox{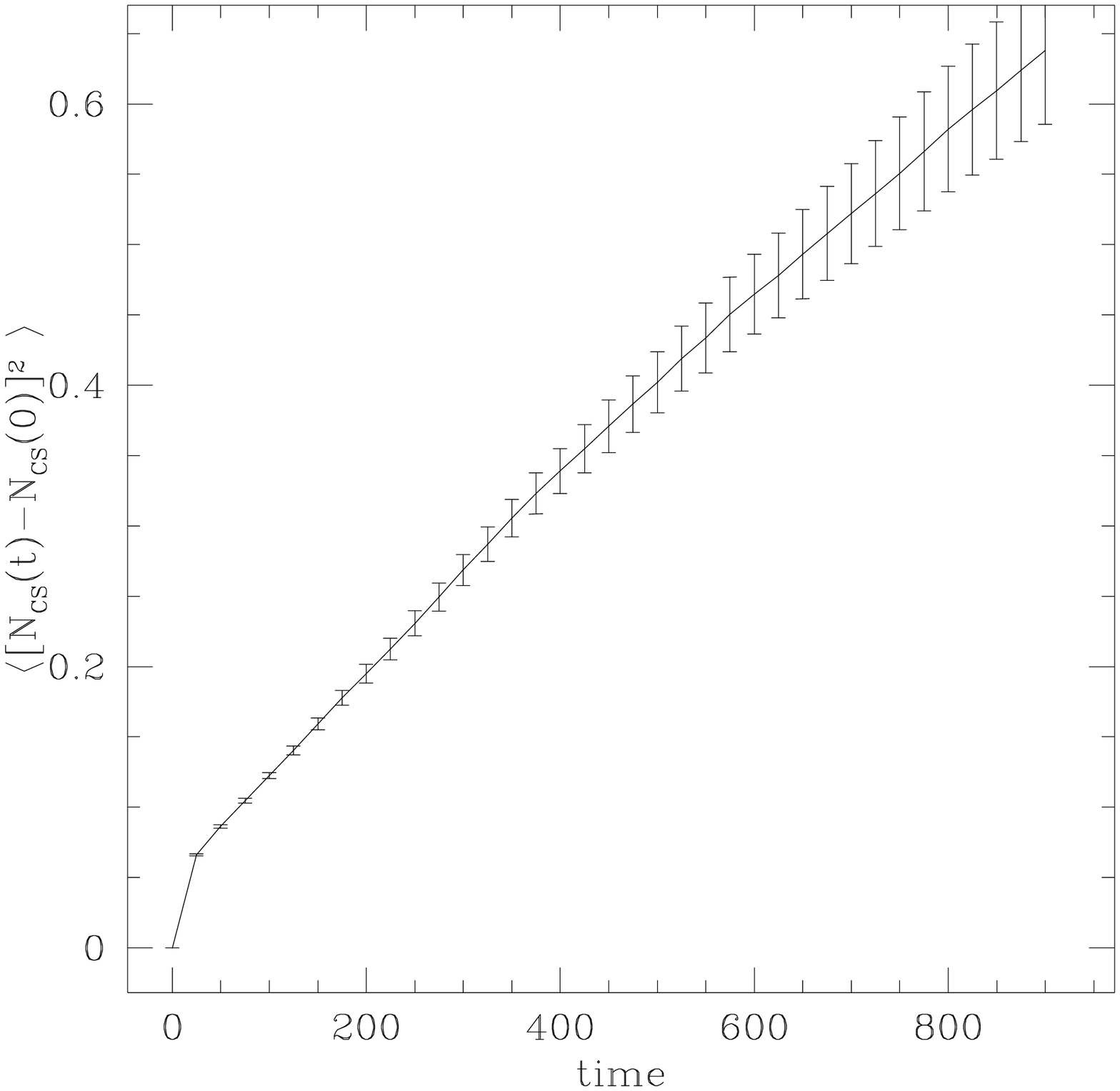}}
\caption{
Same as Fig.~\protect\ref{f4a} after microcanonical averaging 
with $t_b=4150$.
}
\label{f4b}
\end{figure}

The volume in Fig.~\ref{f2} was chosen small
(in addition to large $\btb$) in order to suppress fluctuations 
of different portions of the volume, which would obscure the 
classical vacua.
Fig.~\ref{f3} shows a typical example of the time evolution of 
$Q_L(t)$ for a $24^3$ system at $\btb=13$.
Fig.~\ref{f4a} shows the corresponding average $\langle 
Q^2_L(t)\rangle$ over 33 configurations from the canonical 
ensemble.
Following ref.\ \cite{AmKra95} we shall reduce the errors by 
taking also a microcanonical average, by first averaging over the 
Hamilton part of the trajectories and then over the intitial 
conditions obtained with the Langevin algorithm:
\be
\left[ Q_L^2(t)\right]_{\rm average} =
\left\langle \frac{1}{t_b}\, \int_0^{t_b} dt_0\, \left[
\int_{t_0}^{t_0 + t} dt'\, \sum_{\vecx} q_{\vecx}^L(t')
\right]^2\right\rangle,
\;\;\; 0 < t < t_{\rm run} - t_b.
\label{micro}
\ee
Here the brackets denote the canonical average and $t_{\rm run}$ 
is the maximum time in the run.
The result is in Fig.\ \ref{f4b}. The errors displayed here (and 
elswhere) correspond to the canonical (Langevin) average, the 
microcanonical averages where treated as error free. 

In computing the rate we took care that the diffusion $\langle 
Q^2(t)\rangle$ is large enough to clearly distinguish it from the 
fluctuations about a classical vacuum (technically this is a 
divergence to be removed by subtraction). These fluctations 
can be clearly seen in Fig.\ \ref{f2}, while in Fig.\ \ref{f3} 
they correspond roughly to the width $\approx 0.5$ of the band. 
The autocorrelations of this band produce the upward step near 
$t=0$ in Fig.\ \ref{f4b}, which can be clearly distinguished from 
the subsequent linear increase with $t$. In ref.\ 
\cite{AmKra95} an analytic estimate was made of the contribution 
of these fluctuations to the diffusion in SU(2) gauge theory 
(i.e.\ the magnitude of the step), which roughly fits our data as 
well.

\begin{figure}
\epsfysize 8.5cm
\centerline{\epsfbox{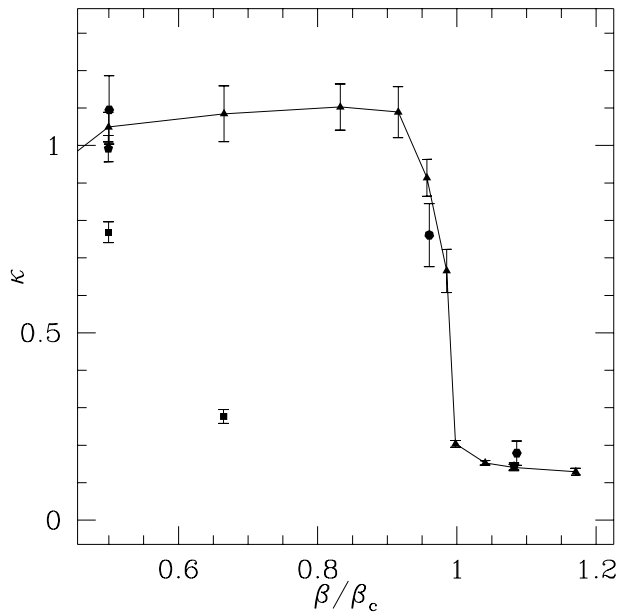}}
\caption{
Results for $\kp \equiv \Gm/L^3 (\al_W T)^4$ as a function of 
$\btb/\btb_c$. The line connects the $N=24$ data.
}
\label{f6}
\end{figure}

The time $t_b$ used for the microcanonical averaging 
(\ref{micro}) was varied from zero, where $t$ could run up to 
$t_{\rm run}$ and the diffusion reached about 3 -- 10 (3.5 in the 
case of Fig.~\ref{f4a}), to values $t_b$ comparable with $t_{\rm 
run}$, where the maximum $t$ had to be much smaller and the 
diffusion reached only values of about 1 ($0.6$ in the case of in 
Fig.~\ref{f4b}). Since such relatively small diffusions 
correspond to only one unit of $\Delta N_{\rm CS}$ one may wonder 
if the diffusion is not dominated by the vacuum fluctuations. 
However, the rates with and without microcanonical averaging were 
consistent, albeit with large errors in the latter case, and 
furthermore, the contribution of vacuum fluctuations can be 
clearly distinguished, as described above (`the step' in 
Fig.~\ref{f4b}).  We used $\btb=13$ as a test case in the broken 
phase and similar tests were made in the symmetric phase were the 
rate is larger. Our conclusion is that a maximum diffusion of 3 
-- 10 without microcanonical average is sufficient for extracting 
the rate, using microcanonical averaging with maximum diffusion 
of order 1, as in Fig.\ \ref{f4b}. This gives us confidence for 
data points with smaller statistics for which microcanonical 
averaging is necessary for a meaningful result.

Using the critical values of $\btb$ and $\vb^2$ as a starting 
point we measured the diffusion $\langle Q^2(t)\rangle$ at 
various temperatures, volumes and lattice spacings. 
We mostly kept $\vb^2$ fixed at $\vb_c^2$ while varying $\btb$,
but we also carried out simulations (for $\btb=13$ and 14) in 
which the deviation of $\vb^2$ away from $\vb^2_c$ was calculated
according to (\ref{VBT}), for comparison.
The results are recorded in Table~\ref{datatable}.
\begin{table}[t]
\centerline{
\begin{tabular}{||l|r|l|c|c|c|r||} \hline
\multicolumn{1}{||c|}{$\bar{\beta}$} &
\multicolumn{1}{c|}{$\bar{\beta}_c$} &
\multicolumn{1}{c|}{N} &
\multicolumn{1}{c|}{$\kappa$} &
\multicolumn{1}{c|}{$\ln [\Gamma/(L^3(\alpha_W T_c)^4)]$} &
\multicolumn{1}{c|}{$T_c/T$} &
\multicolumn{1}{c||}{configs}  \\ \hline
4 & 12 & 24 & 0.81(0.05) & 4.19(0.06) & - & 68 \\ \hline
6 & 12 & 24 & 1.05(0.04) & 2.82(0.04) & - & 88 \\ \hline
8 & 12 & 24 & 1.09(0.08) & 1.71(0.07) & - & 71 \\ \hline
10 & 12 & 24 & 1.10(0.06) & 0.85(0.08) & 0.62 & 80 \\ \hline
11 & 12 & 24 & 1.09(0.07) & 0.48(0.05) & 0.81 & 47 \\ \hline
11.5 & 12 & 24 & 0.92(0.05) & 0.09(0.05) & 0.91 & 36 \\ \hline
11.8 & 12 & 24 & 0.66(0.06) & -0.35(0.09) & 0.96 & 27 \\ \hline
12 & 12 & 24 & 0.204(0.009) & -1.58(0.04) & 1 & 44 \\ \hline
12.5 & 12 & 24 & 0.153(0.006) & -2.04(0.04) & 1.09 & 32 \\ \hline
13 & 12 & 24 & 0.141(0.016) & -2.28(0.11) & 1.18 & 33 \\ \hline
14 & 12 & 24 & 0.127(0.009) & -2.67(0.07) & 1.36 & 7 \\ \hline
4 & 12 & 8 &  0.79(0.05) & 4.15(0.06) & - & 327 \\ \hline
6 & 12 & 8 & 0.77(0.03) & 2.51(0.06) & - & 327 \\ \hline
8 & 12 & 8 & 0.276(0.018) & 0.41(0.13) & - & 93 \\ \hline
6 & 12 & 20 & 1.01(0.03) & 2.78(0.03) & - & 39 \\ \hline
13 & 12 & 20 & 0.145(0.011) & -2.25(0.08) & 1.18 & 28 \\ \hline
10 & 20 & 32 & 1.09(0.09) & 2.86(0.08) & - & 30 \\ \hline
21.7 & 20 & 32 & 0.171(0.017) & -2.08(0.10) & 1.29 & 13 \\ \hline
11.5 & 12 & 32 & 0.98(0.06) & 0.14(0.07) & 0.91 & 10 \\ \hline\hline
13 & 12 & 20 & 0.163(0.018) & -2.13(0.11) & 1.08 & 17 \\ \hline
14 & 12 & 24 & 0.126(0.010) & -2.69(0.08) & 1.17 & 7 \\ \hline
\end{tabular}}
\caption{
Results for the rate.
Here $\vb^2 = \vb^2(\btb_c,N)$ as in Table~\ref{v2table}, 
except for the two bottom lines, where 
$\bar{v}^2$ is related to $\bar{v}_c^2(\btb_c,N)$ 
according to eq.\ (\ref{VBT}), keeping $am_H$ fixed. In the fifth
column $\al_W T_c\equiv 1/a\pi\btb_c $.
}
\label{datatable}
\end{table}
Fig.~\ref{f6} summarizes the results for $\kp \equiv 
\Gm/L^3 (\al_W T)^4$, using $\btb_c$ to set the scale for the 
inverse temperature.
Coming from the lower temperature region we see a jump in the 
rate at the critical temperature, after which 
$\kp$ is roughly constant in the high temperature 
phase, $\kp\approx 1$.
We should be prepared for the possibility that the flatness 
of $\kp$ in the high temperature phase may be partly due to the 
neglect of the renormalization factor $\kp_q$, which is expected 
to increase rapidly with decreasing $\btb$.
The falling of the rate at lower $\btb < 6$ is presumably 
mainly due to the neglect of this finite renormalization $\kp_q$. 
However, there will also be other lattice artefacts in this 
region due to $aT$ getting large ($aT=9/\btb$).

The data in Fig.~\ref{f6} contains tests for volume dependence 
($N=8,20,24$, $\btb_c=12$). We see that the $N \geq 20$ results are 
volume independent within errors, whereas the data for $N=8$ show clear 
finite size effects. The test for the dependence on the lattice 
spacing is less clear cut. We collected data with $\vb^2 = 
\vb_c^2$ corresponding to $(\btb_c,N)=(12,20)$ and (20,32), with 
$\btb/\btb_c = 6/12$ and 13/12 (i.e.\ 
$\btb = 6$, 13 for $\btb_c=12$ and $\btb=10$, 21.7 for 
$\btb_c=20$). The ratio $12/20$ of $\btb_c$ values corresponds to 
a ratio $(12/20)^4 \approx 0.13$ in the lattice rate per unit 
volume, $a\Gm/N^3 = a^4\Gm/L^3$, which is indeed the behavior of the
data, approximately. Plotting the dimensionless $\kp$ at the 
same $\btb/\btb_c$ values we see lattice spacing independence 
within the errors. However, this could be misleading because
the left out $\kp_p$ renormalization factor is different for the 
two lattice spacings. 
 
\begin{figure}
\epsfysize 8.5cm
\centerline{\epsfbox{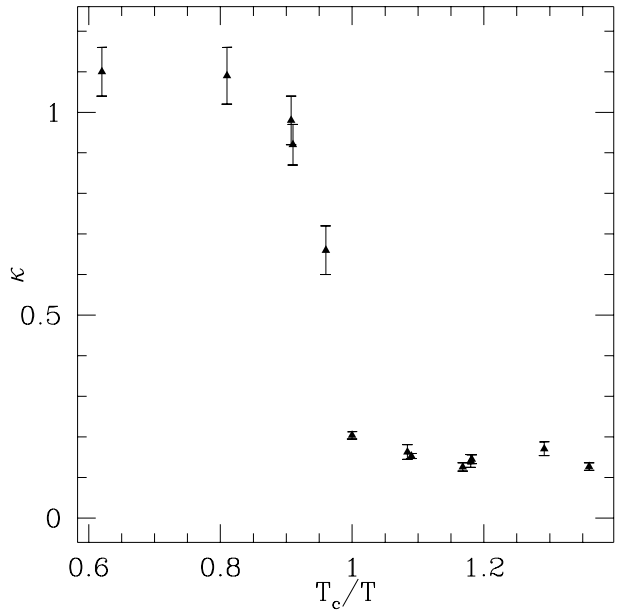}}
\caption{
Plot of $\kp$ as a function of $T_c/T$.
}
\label{f6a}
\end{figure}

There is another aspect that is relevant here. The 
lattice spacing dependence test should be done for a fixed 
physical situation, e.g.\ fixed $T_c/T$. But using $\btb/\btb_c$ 
for the temperature scale is not the same as using $T_c/T$, for 
the data taken at constant $\vb = \vb_c$, since this assumes that 
the lattice spacing $a$ is constant. Changing $\btb$ at constant 
$\vb$ implies changing $a$ by eq.\ (\ref{VBT}) (neglecting the 
running of $m_H$ with $T$). We can try to calculate $T_c/T$ by 
calculating $m_H/T$ for each $\btb - \vb$ pair from (\ref{VBT}) 
and multiplying this by $T_c/m_H = 2.15$ (from Table~\ref{v2table}). We have 
indicated these $T_c/T$ values in Table~\ref{datatable} for the larger volumes 
(the perturbative formula (\ref{VBT}) assumes infinite volume). 
The smaller $\btb$ lead to nonsensical (even imaginary) 
$T_c/T$, since $aT$ gets too large (recall that $aT \geq 0.75$ 
for $\btb\leq 12$). 

Luckily, the constancy of $\kp$ in this high 
temperature region renders the precise shift in $T_c/T$ relative 
to $\btb/\btb_c$ irrelevant. 
For the low temperature point 
the shift in the calculated 
values of $T_c/T = 1.18$ ($\btb=13$) and 1.29 ($\btb=21.7$) 
away from $\btb/\btb_c=13/12 \approx 1.08$
suggest a slightly larger $a$-dependence.
The two data points for which $\vb$ was calculated using (\ref{VBT}) 
(cf.\ the two bottom lines in Table~\ref{datatable}), 
such that $T_c/T = \btb/\btb_c$, are also consistent within errors.
However, the statistics is not sufficient to draw a 
precise conclusion. 
Plotting the data as a function of $T_c/T$ reveals a
$\kp$ which is remarkably constant also in the low temperature phase
(Fig.~\ref{f6a}), rather unlike the sphaleron rate. We come back to this
in the next section.

\begin{figure}
\epsfysize 8.5cm
\centerline{\epsfbox{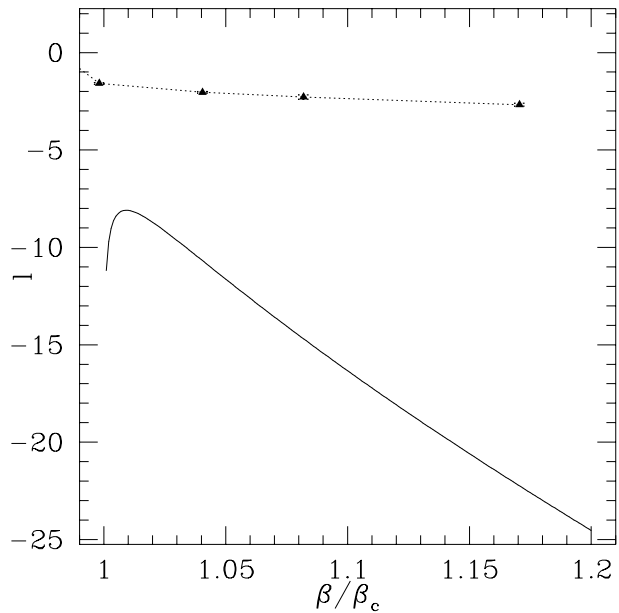}}
\caption{
Comparison of our results for $l\equiv\ln[\Gamma/V(\alpha_W T_c)^4]$ 
versus $\btb/\btb_c$ in 
the broken phase with the sphaleron rate \protect 
(\ref{kappa},\ref{Gan}).
}
\label{f7}
\end{figure}

\section{Conclusion}

We have used dimensional reduction results to pin down the parameters
in the effective hamiltonian for the classical real time simulations. 
The difference between the effective
classical partition function and a more sophisticated dimensional
reduction version appears to be numerically insignificant, at least
for the critical temperature.  We conclude
therefore that the critical temperature is roughly the same as in the
DR work in ref.\ \cite{Faea2} (where $m_H \approx m_W$ as in our
case), $T_c \approx 2.15\; m_H$.
We tested the rate for lattice spacing dependence in the broken 
as well as in the symmetric phase and found it to be roughly 
independent of $a$. We take this as evidence that the important 
parameters in the effective 
real time hamiltonian can indeed be deduced from the DR action. 
We have not taken into account yet the 
finite renormalization factor $\kp_q$ (cf.\ (\ref{kpq})), which 
may give a substantial correction. The neglect of the 
`renormalization of the velocity of light' $z=z_E\approx z_{\pi}$ 
cf.\ (\ref{z},\ref{zrat}) is probably insignificant. The errors 
in our data are presumably underestimated, but since we are still 
in an explorative stage we feel this is acceptable. 

In the symmetric phase $T > T_c$ the ratio $\kp= \Gm/V(\al_W
T)^4\approx 1$ independent of $T$, close to the value $\approx 1.1$
found in ref.\ \cite{AmKra95} for the pure SU(2) gauge theory.  The
rate drops only by a factor of about five when the temperature falls
below the phase transition. In fact, the numerically obtained rate is
much larger than the analytic expression (\ref{kappa},\ref{Gan})
suggests: the maximal value of the analytic rate is about a factor 650
smaller than our numerical result $\kp\approx 0.20$, for $T$ just
below $T_c$ (cf.\ appendix B).  For comparison we have shown the
numerical and analytical rate in Fig.~\ref{f7}. The slopes of the
numerical rate in the logarithmic plot is also much smaller,
indicating a rather small effective sphaleron mass. The numerical data
still seem to be influenced strongly by the phase transition. We do
not expect, of course, the analytic rate to be reliable near the phase
transition, but the discrepancy is surprisingly large. Note that
taking into account the renormalization factor $\kp_q$ would enhance
the difference.

Another puzzling indication for the discrepancy is the following.  If
the rate in the low temperature phase were given by the analytic
expression, we would most probably not have seen a single transition
in the time span in figure \ref{f2}, for the following reasoning. If
the transitions with $\Delta N_{\rm C} \approx 1$ are as clear as in
Fig.~\ref{f2}, the rate may be calculated as $\Gm = n/t$, where $n$ is
the number of transitions that have occured in time $t$, provided
$n,t\to\infty$. Using $n=3$ and $t=1.4 * 10^6$ gives the rough
estimate $\ln[\Gm/L^3(\alpha_W T_c)^4] = -4.8$, with an perhaps error
of a factor of 2 because $n$ is small. This estimate corresponds to a
point in figure \ref{f7} near the linear extrapolation of the
numerical data to $\btb/\btb_c = 3$, far above the analytical 
curve there.  The temperature in Fig.\ \ref{f2} is rather low, 
$T/T_c \approx T/2m_H \approx 1/3$, and one may doubt the 
applicability of dimensional reduction in this case. However, the 
analytic calculation is also based on dimensional reduction and 
it makes sense to make the comparison with the numerical results.

Since the dependence of the rate on the sphaleron energy $E_s$ is
exponential, lattice artefacts in this quantity could give large
systematic errors.  Recent calculations of $E_s$ \cite{VBaPe} have
shown that the difference $\Delta E_s = E_s(a) - E_s(0)$ ($a=$ lattice
spacing) is negative, which would enhance the rate on the
lattice. Although we did not find a significant lattice spacing
dependence near the phase transition,
it is of interest to estimate the systematic error. Ref.\ 
\cite{VBaPe} found that $\Delta B \equiv \al_W \Delta E_s/2m_W 
\approx c\, a^2 m_W^2$, with $c=-0.12$.
Identifying this $m_W$ with the temperature dependent
screening mass $m_W(T)$, we can estimate $\exp(-\Delta E_s/T) 
\approx \exp[|c| a^2 m_W^2(T) (2m_W(T)/T) (1/\al_W)]$.
In the broken phase $m_W(T)$ decreases roughly from $T_c/2$ at 
$T=0$ to $T_c/3$ at $T=T_c$ \cite{Kajea95}. It follows that 
$2m_W(T)/T < 2$ in the region $T_c/2 < T < T_c$, and with 
$aT_c = 3/4$ for $\btb_c=12$, we also have $a^2 m_W^2(T) < a^2 
T_c^2/4 = 9/64$. Putting things together this gives $\exp(-\Delta 
E_s/T) < 2.6$ in the region $T_c/2 < T < T_c$, which amply 
includes our data. Such a large systematic effect on the rate 
would be a substantial error indeed, but it cannot explain a 
factor $ \geq 650$.

Assuming our results to hold up to future scrutiny, they suggest 
that many non-sphaleron processes are contributing to the rate in 
the broken phase. A classification of such configurations is 
given in \cite{Niea95}. The danger of erasure of a surplus of 
baryon number after the electroweak phase transition in the early 
universe, which is already a possibility with the usual sphaleron 
rate, is of course only magnified by these results.\\

\noindent{\bf Acknowledgement}\\
The numerical simulations were carried out
with the code \cite{Kra95} kindly provided to us by Alex 
Krasnitz. We furthermore thank him for stimulating conversations. 
We also would like to thank Pierre van Baal and Margarita 
Garc\'{\i}a P\'erez for useful discussions.
This work is supported by 
FOM and 
NCF, with financial support from 
NWO.

\appendix
\section{Canonical formalism for lattice gauge fields}
For completeness we record here some details on the canonical 
formalism with lattice gauge fields using continuous time 
\cite{KoSu}. Our main objective is to understand the measure 
$\prod (dU dE)$ used in the classical partition function 
(\ref{8}) in relation to the canonical measure $\prod (dA d\Pi)$. 
As an example we consider a one-link lagrangian of the form
\be
L=\frac{1}{zg^2}\, \Tr \dot{U} \dot{U}^{\dagger}
= \frac{1}{2zg^2}\, G_{\al\bt}(A) \dot{A}^{\al}\dot{A}^{\bt},
\ee
where $U$ is a group element with coordinates $A^{\al}$ and 
$G_{\al\bt} = 2\Tr[(\partial U/\partial A_{\al})(\partial 
U^{\dagger}/\partial A^{\bt}]$ is a metric on group space. The 
canonical conjugate to $A^{\al}$ is given by
\be
\Pi_{\al} = \frac{1}{zg^2}\, G_{\al\bt} \dot{A}^{\bt},
\ee
and the hamiltonian is
\be
H = \half\,zg^2 G^{\al\bt} \Pi_{\al} \Pi_{\bt},
\ee
where $G^{\al\bt}$ is the inverse of $G_{\al\bt}$: $G_{\al\bt} 
G^{\bt\gm} = \dl^{\gm}_{\al}$. In the quantization of the system 
it is convenient to introduce left and right handed vielbeins 
according to
\bea
V_{\al}(R) &=& -iU^{\dagger}\frac{\partial U}{\partial A^{\al}}
= V_{\al\bt}(R)\, t_{\bt},\\
V_{\al}(L) &=& iU\frac{\partial U^{\dagger}}{\partial A^{\al}}
= V_{\al\bt}(L)\, t_{\bt},
\eea
where the $t_{\al}$ are generators in the fundamental 
representation of the gauge group. We assume these to be 
normalized as $\Tr t_{\al} t_{\bt} = \dl_{\al\bt}/2$ (for 
SU(2) $t_{\al} = \ta_{\al}/2$). In terms of the $V_{\al\bt}$ and 
their inverse $V^{\al}_{\mbox{ }\bt}$ (i.e.\ $V_{\al\bt} 
V^{\gm}_{\mbox{ }\bt} = \dl^{\gm}_{\al}$ for $R$ and
$L$, respectively), one introduces the $E$'s as
\be
E_{\al}(R) = V^{\bt}_{\mbox{ }\al}(R) \Pi_{\bt},
\ee
and similar for $L$. The $E_{\al}(L)$ and $E_{\al}(R)$ are the 
generators of left and right translations in the sense that
\be
[E_{\al}(L),U] = t_{\al} U,\;\;\;
[E_{\al}(R),U] = U t_{\al},
\ee
follow from the canonical commutation relations 
$[A^{\al},\Pi_{\bt}] = i\dl^{\al}_{\bt}$. Since $G^{\al\bt} = 
V^{\al}_{\mbox{ }\gm}(L) V^{\bt}_{\mbox{ }\gm}(L) = 
V^{\al}_{\mbox{ }\gm}(R) V^{\bt}_{\mbox{ }\gm}(R)$, the 
hamiltonian can be rewritten as
\be
H = \half\, zg^2 E_{\al}(L) E_{\al}(L) = 
\half\, zg^2 E_{\al}(R) E_{\al}(R),
\ee
which provides an attractive prescription for the operator 
ordering problem in the quantum theory. 

Returning to the classical theory, the canonical partition 
function $Z_c$ is given by the integral of $\exp(-H/T)$ over 
phase space with volume element $\prod_{\al} (dA^{\al} 
d\Pi_{\al})$. As usual, this volume element is invariant under 
transformations of coordinates $A^{\al}$. It is, 
however, generally more transparant to trade the $\Pi$'s for the 
$E$'s. Then the invariant volume element on group 
space $dU$ arises naturally from $\Pi_{\al} = V_{\al\bt}(L) 
E_{\bt}(L)$,
\bea
\prod_{\al} (dA^{\al} d\Pi_{\al})
&=& |\det V(L)| \prod_{\al} [dA^{\al} dE_{\al}(L)]
= \sqrt{\det G }\, \prod_{\al} [dA^{\al} dE_{\al}(L)] \nonumber\\
&=& {\rm const.}\; dU \,\prod_{\al} dE_{\al}(L),
\eea
and similar for $R$.

The conventions used in ref.\ \cite{Kra95} can be reached from 
those in sect.~3
by first absorbing $z$ in the time scale and then writing $L = 
(4/g^2) \bar{L}$. This $\bar{L}= (1/4) \Tr \dot{U} 
\dot{U}^{\dagger}$ is equal to $L$ with $z=4/g^2$ and going 
through the canonical formalism again gives $\bar{H} = 2E^2$ in 
terms of the new canonical $E$'s. The $\bar{E}_{\al}$ are related 
to these by $\bar{E}_{\al} = 2E_{\al}$.

\section{Analytic rate}
We give here some details on the analytic form 
(\ref{kappa},\ref{Gan}) for the rate. The function 
$f(\lm_3/g^2_3)$ in (\ref{Gan}) is given by
\be
f = \frac{\rh}{B^7}\, \frac{{\cal N}_{\rm tr} {\cal N}_{\rm 
rot}}{16\pi^2}\, \dl
\ee
where $\rh = |\om_-|/2m_W$ with $\om_-$ the unstable sphaleron 
eigenvalue, $B=\al_W E_s/2m_W$, ${\cal N}_{\rm tr}$ and ${\cal 
N}_{\rm rot}$ are zero mode factors and $\dl$ is the ratio of 
fluctuation determinants. We treat the temperature dependence of 
various quantities in a customary approximation, in which $\rh$, 
$B$, ${\cal N}_{\rm tr}$, ${\cal N}_{\rm rot}$ and $\dl$ are
independent of $T$ and depend only on $\lm_3/g^2_3 \approx 
8m_H^2/m_W^2$. All temperature dependence is then in $m_W$ which 
is approximated as
\be
m_W(T) = m_W(0)\, \sqrt{1-T^2/T_c^2}.
\ee
For the case $m_H = m_W$ we have $\rh \approx 0.73$, $B\approx 
1.8$, ${\cal N}_{\rm tr} \approx 7$, ${\cal N}_{\rm rot}\approx 
12$ \cite{Caea90} and $\dl\approx \exp(-9.64)$ \cite{BaJu94}. The 
temperature dependence in eq.\ (\ref{kappa}) now enters through 
the factor $x^7\exp(-x)$, $x= 2B m_W(T)/\al_W T$, which has a 
maximum of $\approx 751$ at $x=7$, or $T/T_c \approx 0.99$. The 
rate at this maximum is given by $\kp \approx 3.1 \times 
10^{-4}$, which is a factor of about 650 smaller than the 
value $\approx 0.20$ found numerically for $T$ just below the 
phase transition.


\begin{thebibliography}{99}
\bibitem{CoKaNe93} A.G. Cohen, D.B. Kaplan, A.E.~Nelson,
                   Ann.\ Rev.\ Nucl.\ Part.\ Sci.\ 43 (1993) 27.
\bibitem{RuSha96}  V.A.~Rubakov and M.E.~Shaposhnikov, hep-ph/9603208.
\bibitem{KleSha}   S.~Yu.~Klebnikov and M.E.~Shaposhnikov,
                   Nucl.\ Phys.\ B308 (1988) 885.
\bibitem{ArMcLe87} P.~Arnold and L.~McLerran, Phys.\ Rev.\ D36 (1987) 581.
\bibitem{Caea90}   L.~Carson, X.~Li, L.~McLerran and R.~Wang,
                   Phys.\ Rev.\ D42 (1990) 2127.
\bibitem{BaJu94}   J.~Baacke and S.~Junker, Phys.\ Rev.\ D49 (1994) 2055;
                   (E) Phys.\ Rev.\ D50 (1994) 4227.
\bibitem{Moo95}    G.~Moore, Phys.\ Rev.\ D53 (1996) 5906.
\bibitem{Diaea94}  D.~Diakonov, M.~Polyakov, S.~Sieber, J.~Schaldach and
                   K.~Goeke, Phys.\ Rev.\ D49 (1994) 6964.
\bibitem{Ph95}     O.~Philipsen, Phys.\ Lett.\ B358 (1995) 210.
\bibitem{GrRu}     D.Yu.~Grigoriev and V.A.~Rubakov, Nucl.\ Phys.\ B299 (1988) 67.
\bibitem{GrRuSha}  D.Yu.~Grigoriev and V.A.~Rubakov and M.E.~Shaposhnikov,
                   Nucl.\ Phys.\ B326 (1989) 737.
\bibitem{BochFo93} A.~Bochkarev and P.~de Forcrand, Phys.\ Rev.\ D47 (1993) 3476.
\bibitem{KraPo94}  A.~Krasnitz and R.~Potting, 
                   Nucl.\ Phys.\ B (Proc.\ Suppl.) 34 (1994) 613.
\bibitem{SmTa94}   J.~Smit and W.H.~Tang, Nucl.\ Phys.\ B (Proc.\ Suppl.) 34 (1994) 616.
\bibitem{FoKraPo94}P.~de Forcrand, A.~Krasnitz and R.~Potting, 
                   Phys.\ Rev.\ D50 (1994) 6054.
\bibitem{SmTa95}   J.~Smit and W.H.~Tang, Nucl.\ Phys.\ B (Proc.\ Suppl.) 42 (1995) 590.
\bibitem{Ambea}    J.~Ambj$\o$rn, T.~Askgaard, H.~Porter and M.E.~Shaposhnikov,
                   Phys.\ Lett.\ B244 (1990) 479; Nucl.\ Phys.\ B353 (1991) 346.
\bibitem{Kra95}    A.~Krasnitz, Nucl.\ Phys.\ B455 (1995) 320.
\bibitem{Moo96}    G.~Moore, hep-ph/9603384. 
\bibitem{AmKra95}  J.~Ambj\o rn and A.~Krasnitz,  Phys.\ Lett.\ B362 (1995) 97. 
\bibitem{Faea1}    K.~Farakos, K.~Kajantie, K.~Rummukainen  and M.~Shaposhnikov,
                   Nucl.\ Phys.\ B442 (1995) 317.
\bibitem{Faea2}    K.~Farakos, K.~Kajantie, K.~Rummukainen and M.~Shaposhnikov, 
                   Phys.\ Lett.\ B336 (1994) 494.
\bibitem{Kaea96}   K.~Kajantie, M.~Laine, K.~Rummukainen and M.~Shaposhnikov,
                   Nucl.\ Phys.\ B458 (1996) 90.
\bibitem{RG}       A.~Hasenfratz, P.~Hasenfratz, U.~Heller and F.~Karsch,
                   Phys.\ Lett.\ B143 (1984) 193. R.~Gupta and A.~Patel, 
                   Nucl.\ Phys.\ B251 (1985) 789.
\bibitem{KeMaPa}   U.~Kerres, G.~Mack, G.~Palma, hep-lat/9505008.
\bibitem{JaPaPe96} A.~Jakov\'ac, A.~Patk\'os and P.~Petreczky,
                   Phys.\ Lett.\ B367 (1996) 283. 
\bibitem{BoMcLeSmil95}  D.~Bodeker, L.~McLerran, A.~Smilga, 
                   Phys.\ Rev.\ D52 (1995) 4675.
\bibitem{Boch93}   A.~Bochkarev, Phys.\ Rev.\ D48 (1993) 2382.
\bibitem{Kro}      A.S.~Kronfeld, Nucl.\ Phys.\ B (Proc.\ Suppl.) 4 (1988) 329.
\bibitem{DiGia}    B.~All\'es, M.~Campostrini, L.~Del Debbio, A.~ Di Giacomo,
                   H.~ Panagopoulos, E.~Vicari, Phys. Lett. B 336 (1994) 248.
\bibitem{SmVi}     J.~Smit and J.C.~Vink, Nucl.\ Phys.\ B284 ( 1987) 234;
                   Nucl.\ Phys.\ B298 (1988) 557.
\bibitem{Vi}       J.C.~Vink, Phys.\ Lett. B212 (1988) 483. 
\bibitem{AmbFaHaKouThor} J.~Ambj$\o$rn, K.~Farakos, S.~Hands, G.~Koutsoumbas
                   and G.~Thorleifsson, Nucl.\ Phys.\ B245 (1994) 39.
\bibitem{KoSu}     J.B.~Kogut and L.~Susskind, Phys.\ Rev.\ D11 (1975) 395.
\bibitem{VBaPe}    M.G.~P\'erez and P.~van Baal, hep-lat/9512004.
\bibitem{Kajea95}  K. Kajantie, M. Laine, K. Rummukainen and M. Shaposhnikov,
                   hep-lat/9510020.
\bibitem{Niea95}   M.~Axenides, A.~Johansen and H.B.~Nielsen, hep-ph/9511240. 
\end{thebibliography}
\end{document}